# Antiferromagnetic Order of the Iron Spins in NdOFeAs

Ying Chen[1,2], J. W. Lynn[1], J. Li[1,2], G. Li[3], G. F. Chen[3], J. L. Luo[3], N. L. Wang[3], Pengcheng Dai[4,5], C. dela Cruz[4,5], and H. A. Mook[5]

[1]NIST Center for Neutron Research, National Institute of Standards and Technology, Gaithersburg, Maryland 20899-6102
[2]Department of Materials Science and Engineering, University of Maryland, College Park, Maryland 20742-6393
[3]Beijing National Laboratory for Condensed Matter Physics, Institute of Physics, Chinese Academy of Sciences, Beijing 100190, China
[4]Department of Physics and Astronomy, The University of Tennessee, Knoxville, Tennessee 37996-1200
[5]Neutron Scattering Science Division, Oak Ridge National Laboratory, Oak Ridge, Tennessee 37831

Abstract

Polarized and unpolarized neutron diffraction measurements have been carried out to investigate the iron magnetic order in undoped NdOFeAs. Antiferromagnetic order is observed below 141(6) K, which is in close proximity to the structural distortion observed in this material. The magnetic structure consists of chains of parallel spins that are arranged antiparallel between chains, which is the same in-plane spin arrangement as observed in all the other iron oxypnictide materials. Nearest-neighbor spins along the *c*-axis are antiparallel like LaOFeAs. The ordered moment is 0.25(7) $\mu_B$, which is the smallest moment found so far in these systems.

## I. Introduction

The nature of the magnetic order in superconductors has had a rich and interesting history, and has been a special topic of interest ever since the parent materials of the high $T_C$ cuprates were found to be antiferromagnetic Mott insulators that exhibit huge exchange energies within the Cu-O planes. For the newly discovered iron oxypnictide class of superconductors,[1-18] the observation of long range spin density wave antiferromagnetic order in the undoped materials has naturally led to strong parallels being drawn between these two classes of materials.[19-29] The magnetic structure of the oxypnictides within the *a-b* plane consists of chains of parallel Fe spins that are coupled antiferromagnetically in the orthogonal direction, with an ordered moment substantially less than one Bohr magneton. Hence these are itinerant electron magnets, with a spin structure that is consistent with Fermi-surface nesting along with possible strong electron correlation effects.[30-45] Here we report the observation of antiferromagnetic order in the parent compound of one of the highest $T_C$ systems, NdOFeAs. The magnetic structure is the same as that for LaOFeAs, but with an ordered moment of only 0.25 $\mu_B$ which is the smallest observed so far in this class of materials.

## II. Experimental Procedures

The neutron diffraction measurements were carried out with the BT-7 spectrometer at the NIST Center for Neutron Research, using the diffraction mode with a position sensitive detector (PSD) that covered an angular range of approximately five degrees. The neutron wavelength employed was 2.359 Å using a pyrolytic graphite (PG) monochromator, and PG filter to suppress higher-order reflections to achieve a monochromatic incident beam. The diffraction patterns were collected with a 50′ full-width-at-half-maximum Söller collimator before the sample, and an 80′ radial collimator between the sample and the PSD. Data were obtained in steps of 0.25° so that the intensity at each scattering angle was measured many times, and then the data were binned to obtain the diffraction pattern. Measurements of the magnetic scattering versus temperature were obtained at a fixed angular position and varying the temperature. The polycrystalline sample weighed approximately 8.2 g and was prepared using the method described elsewhere [20]. The sample was sealed in an aluminum container with helium exchange gas and mounted on the cold finger of a closed cycle helium refrigerator. Data were collected in the temperature range from 5.5 K to 225 K.

### III. Experimental Results

Figure 1 shows the diffraction data taken at 30 K, which is well above the ordering temperature of Nd in order to avoid possible influence of the Nd moments on the scattering. The Nd spins order around 2 K, and this ordering has already been investigated by powder magnetic diffraction, while the iron magnetic order was too weak to be observed [46,47]. In the present work three magnetic peaks originating from the iron spins have been observed between the strong structural Bragg peaks. The intensities of these peaks are considerably weaker than observed in the other materials investigated so far [19, 22, 23] and indicate that the ordered moment is smaller (for example compare with Fig. 3 of ref. 19). The peaks can be indexed on the basis of the magnetic unit cell that consists of chains of parallel spins along one of the in-plane axes, and antiparallel along the other. This is the identical spin configuration as found for all the other materials investigated so far. The c-axis, on the other hand, needs to be doubled to describe the magnetic structure, which means that nearest-neighbor spins are antiparallel in that direction. This is the same magnetic structure as observed in LaOFeAs [19], which is shown in Fig. 2, with an ordered moment of 0.25(7).

The temperature dependence of the magnetic intensity of the (1,0,3) magnetic peak is shown in Fig. 3. The solid curve is a fit to mean-field theory and gives an estimated ordering temperature of $T_N$=141(6) K. The uncertainty here represents one standard statistical deviation, and is the uncertainty indicated throughout this work when error bars are presented. Alternatively, a power law fit to the data using

$$I \propto (T_N - T)^{2\beta} \quad,$$

is shown as the dashed curve. The lower temperature data were excluded from the fit since a power law is not expected to valid well away from the critical temperature. This fit yields an ordering temperature of $T_N = 143(5)$ K and $\beta$=0.25(4). The value of β is smaller than that expected for a three-dimensional Heisenberg system, and likely is a reflection of the layered nature of the magnetic system. We remark that when there are two types of magnetic spins in the system, the spins that order at higher temperatures can induce a moment on the spins that order at lower T, such as is found in the related $Nd_2CuO_4$ material [49,50]. The size of the induced moment is just a reflection of the susceptibility of the Nd, which increases with decreasing T. Thus we have also performed mean field fits excluding the data below 30 K, but the fitted value of the ordering temperature remained within the stated uncertainties. We therefore don't believe that the present fits are significantly influenced by the Nd moments.

For all of these iron-based undoped systems the magnetic scattering from the iron spins is quite weak, and develops near or below the temperature of the structural phase transition. To establish that this scattering is indeed magnetic in origin and not structural scattering associated with the lattice distortion, we carried out polarized neutron diffraction measurements with polarization analysis of the scattered neutrons, using $He^3$ polarizers before and after the sample.[51] The polarization analysis technique as applied to this problem is in principle straightforward [52,53]. Nuclear coherent Bragg scattering never causes a reversal, or spin-flip, of the neutron spin direction upon scattering. Thus ideally the nuclear peaks will only be observed in the non-spin-flip scattering geometry. We denote this configuration as (+ +), where the neutron is incident with up spin, and remains in the up state after scattering. Non-spin-flip scattering also occurs if the incident neutron is in the down state, and remains in the down state after scattering (denoted (− −)). The magnetic cross sections, on the other hand, depend on the relative orientation of the neutron polarization **P** and the reciprocal lattice vector τ. In the configuration where **P**⊥τ, half the magnetic Bragg scattering involves a reversal of the neutron spin (denoted by (− +) or (+ −)), and half does not. Thus for the case of a purely magnetic reflection the spin-flip (− +) and non-spin-flip (+ +) intensities should be equal in intensity. For the case where **P**∥τ, all the magnetic scattering is spin-flip, and should be twice as strong as for the **P**⊥τ configuration, while ideally no non-spin-flip scattering will be observed. Fig. 4 shows the data for the (1,0,3) magnetic Bragg peak and the (0,0,2) nuclear Bragg peak. The instrumental flipping ratio decreased from a high of 17 during the experiment as the $He^3$ cells relaxed, and for the data in the figure the non-spin-flip to spin-flip scattering for the nuclear peak is 10, the instrumental flipping ratio at that time. This indicates that this peak is indeed purely structural in origin. For the magnetic peak, on the other hand, we see that the peak is observed in the spin-flip scattering geometry, indicating that it has a magnetic contribution. For the **P**⊥τ configuration the spin-flip scattering is reduced as expected. Note also that the "background" is lower, which is due to the paramagnetic scattering from the Nd spins; paramagnetic scattering follows the same selection rule as the magnetic Bragg scattering, that the spin-flip scattering for the **P**∥τ geometry is double the intensity for the **P**⊥τ geometry. Then the subtraction of the **P**⊥τ intensity from the **P**∥τ intensity eliminates any structural cross sections. The observed peak, (together with the Nd paramagnetic "background") confirms that the temperature-dependent peak observed in Fig. 3 is magnetic in origin.

We remark that in the course of these measurements we also noted that there is residual scattering in the non-spin-flip channel at the position of this magnetic peak, so there is some very weak structural scattering at this position as well. The unpolarized beam measurements indicate that this scattering is still present at 225 K, and hence it is not associated with either the structural or magnetic transition. The position of this very small structural peak indexes precisely with the lattice of the primary phase, and in particular does not coincide with any known impurity phases. Therefore we attribute it to scattering from NdOFeAs, which indicates a subtle addition to the crystallographic structure, which has not been studied previously at this level of detail. It is likely that single crystals will be necessary to unravel some of these more subtle crystallographic issues in these materials.

In summary, we have observed the magnetic ordering of the iron spins in NdOFeAs, which order at 141 K with the same in-plane spin configuration as found for all the systems investigated so far, and with the same antiparallel *c*-axis arrangement as observed in LaOFeAs. For the cuprate superconductors the parent materials are Mott insulators, where a single electron is localized on the Cu site. In the iron-based family of superconductors, on the other hand, the undoped systems are antiferromagnetically ordered metallic materials, and are better thought of as itinerant electron systems. For the cuprates all the spin structures in the *a-b* plane are simple collinear antiferromagnets where nearest neighbors are antiparallel, while in the present systems the iron spins only order when the crystal distorts, with the spins parallel in one direction and antiparallel in the other, as expected from calculations. It is clear that the Fermi surface is playing an essential role in the magnetic ordering, and thus with the small moment it may be justified to refer to this state as a spin density wave even though it is commensurate in nature. In addition, correlation effects and on-site physics may also play an important role. Upon doping into the superconducting state the crystal distortion and magnetic order vanish in the system that has so far been investigated [22], again suggesting that these two phenomena are closely related. In the tetragonal superconducting state the magnetic interactions are therefore expected to be frustrated, yielding strong spin fluctuations in analogy with the cuprates, and it will be particularly interesting to explore the spin excitations in both the parent and superconducting systems. One of the exciting aspects of these new superconductors is that they belong to a comprehensive class of materials where many chemical substitutions are possible. This versatility is already opening up new research avenues to understand the origin of the superconductivity, and should also enable the superconducting properties to be tailored for commercial technologies. There is no doubt that the newly discovered materials have re-energized the superconductivity community.

### IV. Acknowledgements

We thank Qingzhen Huang and Yasutomo Uemura for helpful discussions. This work is supported in part by the US National Science Foundation through DMR-0756568, and by the US Department of Energy, Division of Materials Science, Basic Energy Sciences, through DOE DE-FG02-05ER46202. The work at the Institute of Physics, Chinese Academy of Sciences, is supported by the National Science Foundation of China, the Chinese Academy of Sciences and the Ministry of Science and Technology of China.

**Figure Captions**

Fig. 1. (color online) A portion of the diffraction pattern taken on the BT-7 triple axis spectrometer is shown, indicating the iron magnetic Bragg peaks and the structural peaks in NdOFeAs. The data were collected at 30 K, well above the ordering of the Nd to avoid any significant contribution from those moments. The structural peaks are labeled using the tetragonal unit cell, while the magnetic peaks are indexed with the magnetic unit cell which is twice as long along the *c*-axis ($2c_N$).

Fig. 2. (color online) The antiferromagnetic spin structure of the iron moments, with moments directed in the *a-b* plane. The moments are indicated as along one of the axes for illustrative purposes, but it was not possible in these measurements to determine the spin direction within the plane since the magnetic intensities were too weak to distinguish *a* from *b*.

Fig. 3 (color online) Temperature dependence of the magnetic intensity of the (1,0,3) peak. The solid curve is a fit to mean field theory to provide an estimate of $T_N$=141(6) K for the ordering temperature. The dashed curve is a power law fit as described in the text. The magnetic intensity is proportional to the square of the ordered (staggered) moment.

Fig. 4. (color online) Polarized neutron diffraction results. a) shows the spin flip (solid (red) circles) and non-spin-flip scattering (solid (green) squares) for the (0,0,2) structural peak at ~32° the and the magnetic peak at ~34°. The ratio of the intensities for the structural peak is just the instrumental flipping ratio, while the magnetic peak is observed in the spin-flip scattering. Error bars are smaller than the data points. b) solid (green) squares are for the **P**∥τ configuration and the solid (green) diamonds are for the **P**⊥τ configuration. The change in the "background" originates from the (diffuse) paramagnetic scattering of the Nd moments. c) The subtraction of the scattering in the **P**⊥τ configuration from the scattering in the **P**∥τ configuration yields the purely magnetic peak (plus the paramagnetic diffuse component).

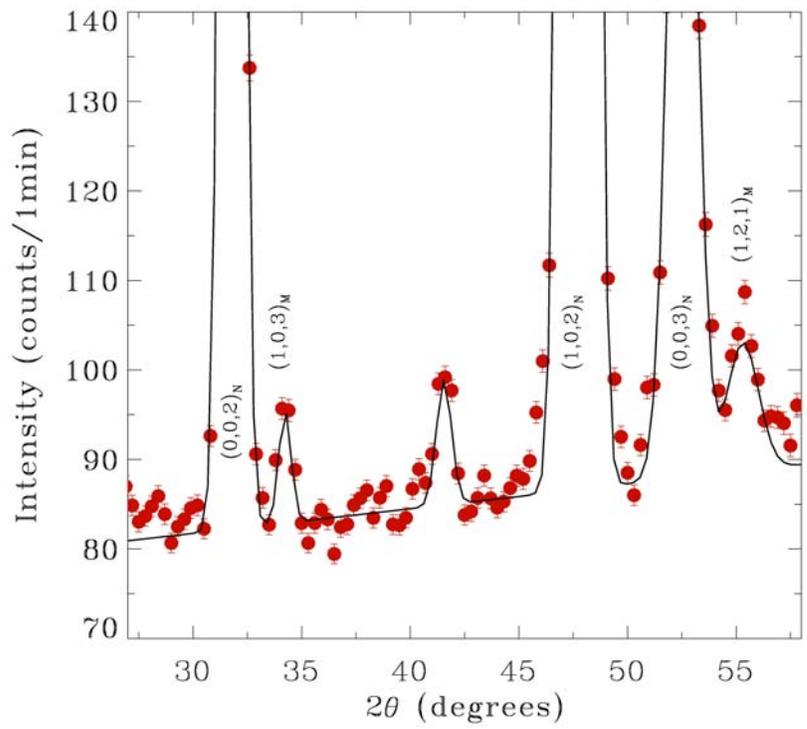

Fig. 1

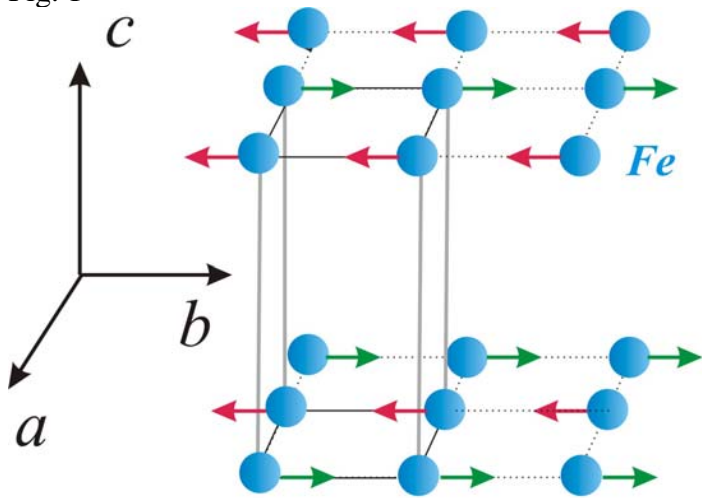

Fig. 2

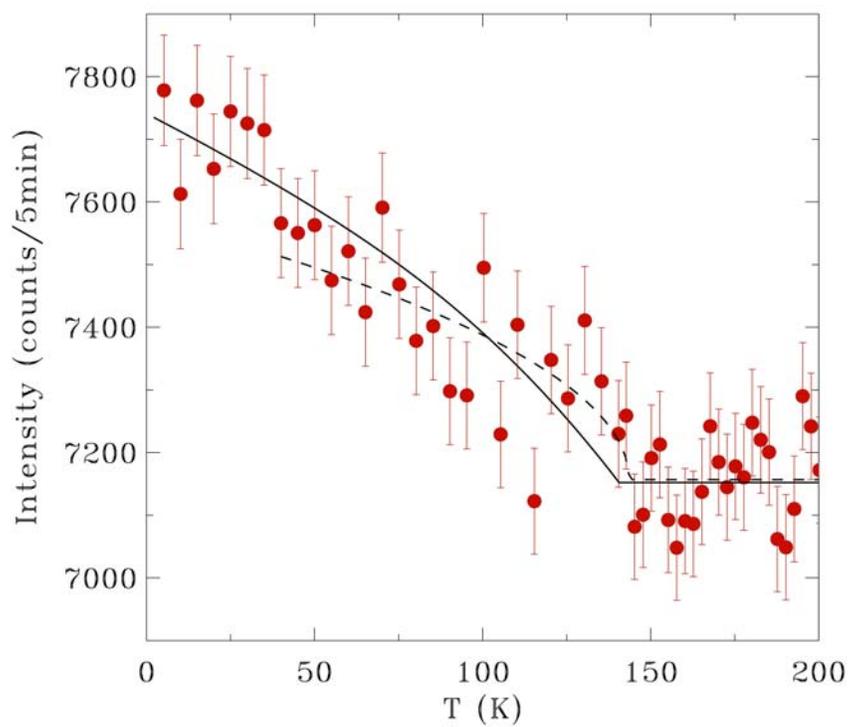

Fig. 3

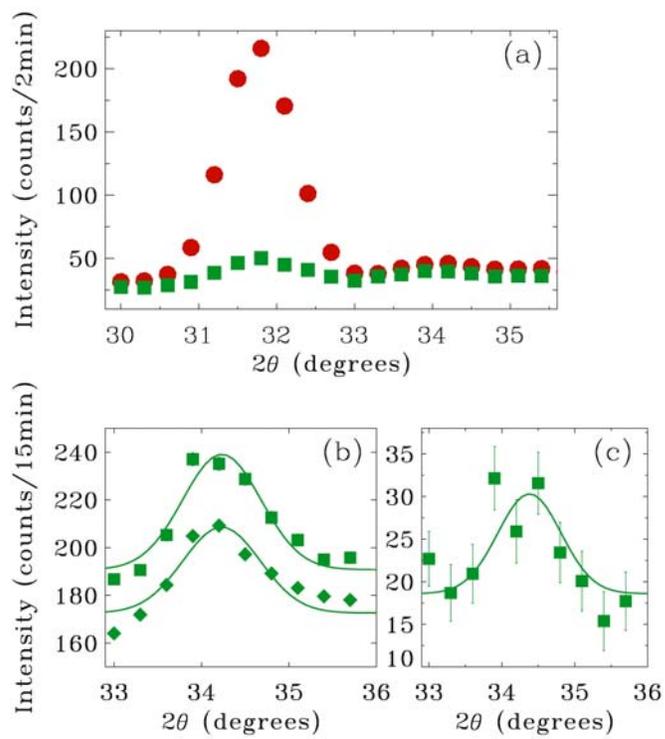

Fig. 4